March 1995 (K)        May 1995 (B)        August 1995 (R)

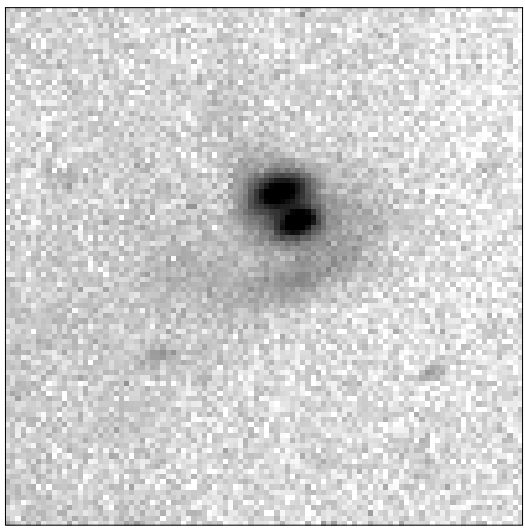 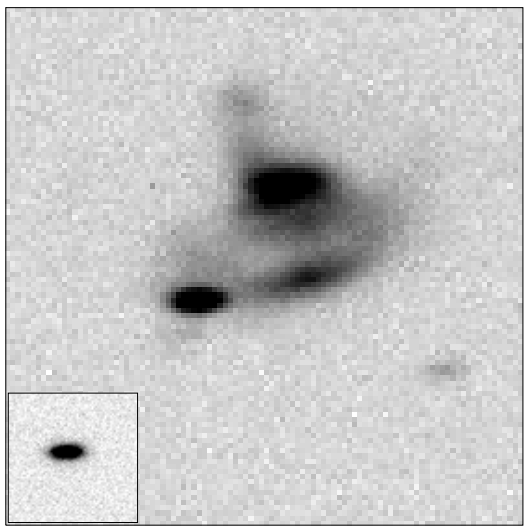 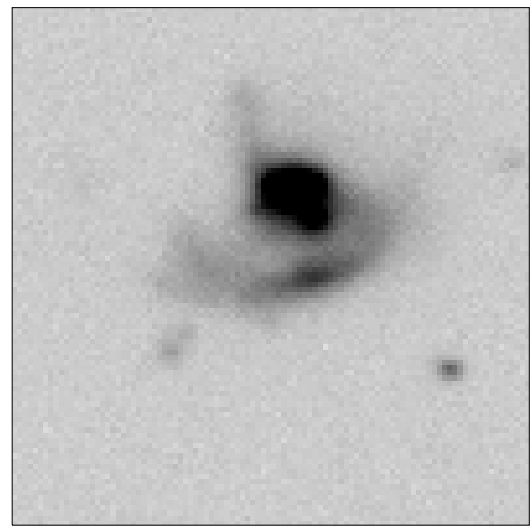

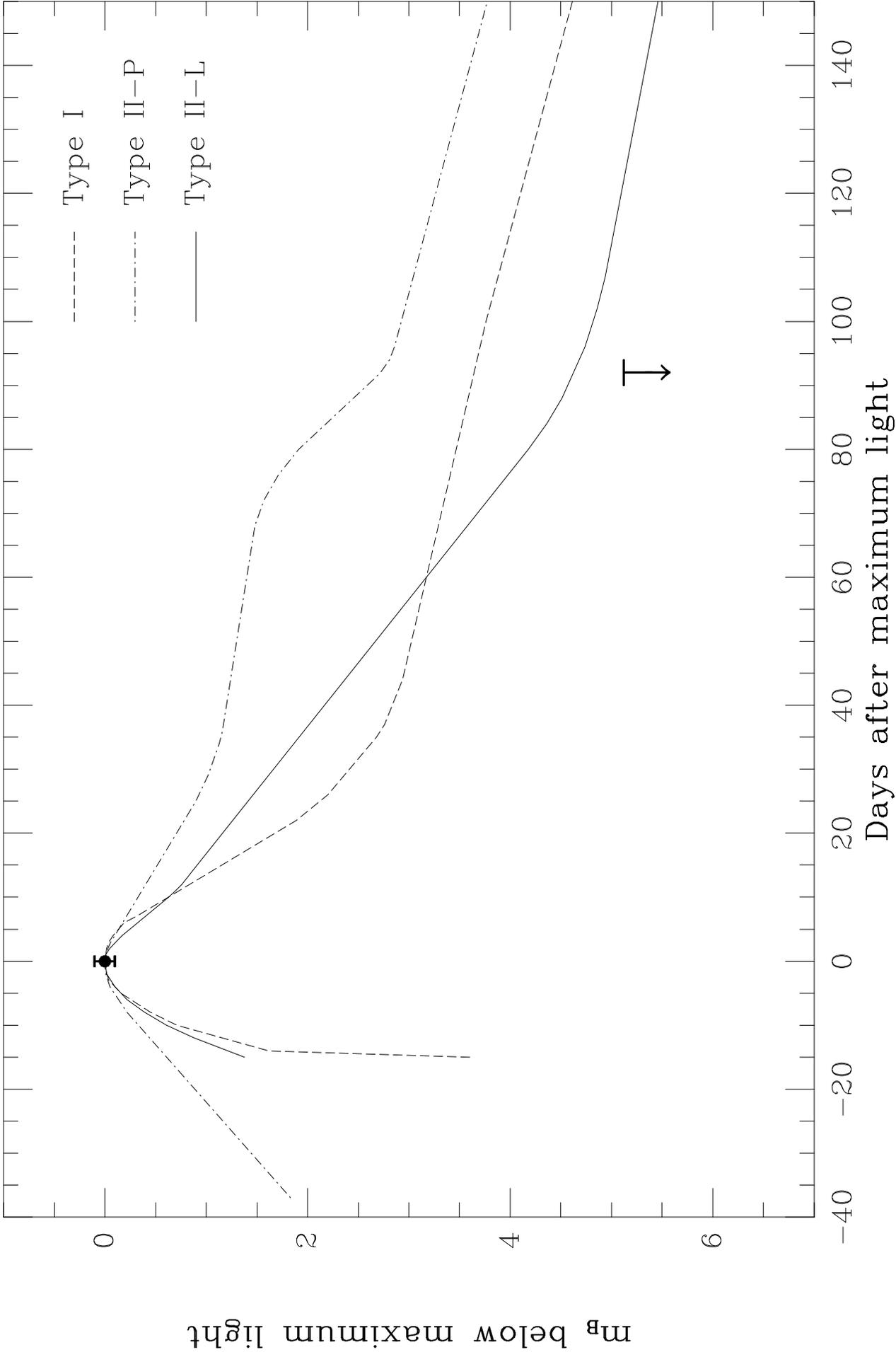

# A candidate supernova in the ultraluminous infrared galaxy IRAS 12112+0305

N. Trentham

Institute for Astronomy
University of Hawaii
Honolulu HI 96822
USA

**Abstract.** We report the discovery of a candidate supernova in the ultraluminous galaxy IRAS 12112+0305 during April/May 1995. This would be the first supernova discovered in an ultraluminous infrared galaxy ($L_{\rm IR} > 10^{12} L_\odot$), although supernovae have been discovered in the slightly less extreme starburst galaxy NGC 3690. If this is a supernova, our very limited observations suggest that it is a Type II.

## 1. Introduction

Ultraluminous infrared galaxies (hereafter ULIGs) are dissipative collapses in progress (see Sanders & Mirabel 1996 for a recent review). Their huge infrared luminosities ($L_{\rm IR} = L_{8-1000\mu} > 10^{12} L_\odot$) are the result of the thermal emission from dust surrounding either starbursts or active galactic nuclei (AGN). Large amounts of gas have fallen to small radii resulting in extremely high densities; this gas is observed through its molecular transitions (Solomon et al. 1992). These high densities are presumably responsible for the starburst or AGN (Sanders et al. 1988) which in turn is reponsible for the huge $L_{\rm IR}$. High-resolution imaging of ULIGs indicates that the usual trigger of this collapse is a recent merger (Clements et al. 1996).

Therefore, we might expect to see a high rate of Type II Supernovae (SN II) in ULIGs. This is because SN II come from the core-collapse explosion of very massive young stars, in contrast to the normally more luminous Type Ia supernovae (hereafter SN Ia) which come from exploding carbon-oxygen white dwarfs. Until now, none have been discovered in the extremely luminous ULIGs with $L_{\rm IR} > 10^{12} L_\odot$; there are $\sim 40$ such galaxies known (Sanders & Mirabel 1996). However two SN (Treffers et al. 1993, van Buren et al. 1994) have been recently observed in the slightly less extreme starburst galaxy NGC 3690 ($L_{\rm IR} = 7.9 \times 10^{11} L_\odot$, Carico et al. 1990). This paper reports the discovery of a candidate supernova (hereafter SN) in the ULIG IRAS 12112+0305 ($L_{IR} = 1.9 \times 10^{12} L_\odot$, Sanders et al. 1988). Our observations are limited but do indeed suggest that this is a SN II.

Throughout this paper we assume $H_0 = 75$ km s$^{-1}$ Mpc$^{-1}$.

## 2. Observations and Photometry

The observations (see Figure 1) presented here were taken as part of a larger study (Trentham et al. 1996); the discovery of this SN was serendipitous. On March 13 1995, IRAS 12112+0305 was imaged for 12 minutes with a $K'$ filter (Wainscoat & Cowie 1992) using the QUIRC 1024 × 1024 HgCdTe array at the f/10 Cassegrain focus of the University of Hawaii 2.24 m telescope on Mauna Kea. On May 1 1995, IRAS 12112+0305 was imaged for 8 minutes with a Mould B filter using a thinned Tektronix 2048 × 2048 CCD at f/10 on the UH 2.24 m telescope. This image revealed a bright point-source in a southern extension of the galaxy that was not visible in the March image. We interpret this point-source as a SN for the reasons detailed at the beginning of the next section. On August 1 1995, we obtained Mould $V$ (5 minutes), $R$ (15 minutes), and $I$ (5 minutes) images of IRAS 12112+0305 with the same observational setup as for the May images. We failed to detect at the $3\sigma$ level any point-source at the position of the bright point-source in the B-band image.

The images were reduced using standard techniques. The photometry was converted to the UBVRI magnitude system of Landolt (1992) for the optical images and the $K'$ system of Wainscoat & Cowie (1992) for the QUIRC image; the zero points are accurate to 2-3%. We used the

*Send offprint requests to*: N. Trentham

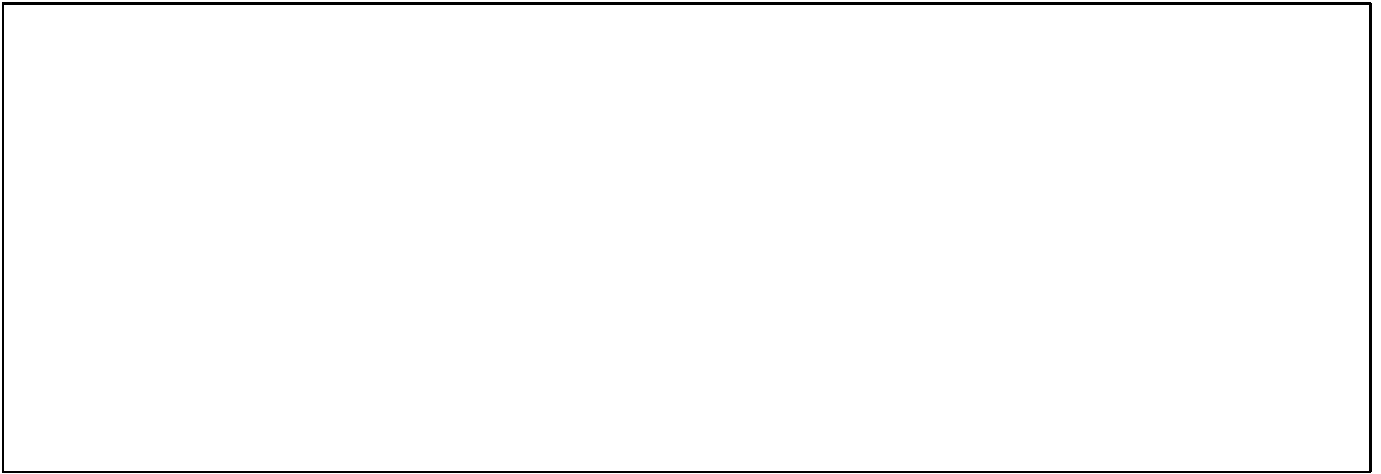

**Fig. 1.** Images of IRAS 12112+0305 taken as described in the text. North is up and East is to the left in all images. Each image is a square of side 41 arcseconds. The SN appears clearly as a bright point source in the May image. The insert in the May image shows a psf star – the image is elongated because of heavy windshake of the telescope.
Note also the secondary nucleus, which is clearly visible in the $K$-band image, but is completely obscured by dust in the galaxy in the $B$-band image.

psf-fitting algorithm DAOPHOT (Stetson 1987) to measure the total apparent magnitude $m_B = 18.83$ of the SN in the May $B$-band image. We estimate an uncertainty in this magnitude of $\sim 0.1^m$. This error is larger than one normally expects from psf-fitting, but the total light contributed by the background galaxy is uncertain. We obtained limiting magnitudes of the SN in the March and August images as follows. First, the sky noise $\sigma_{\rm rms}$ and seeing $b_{\rm FWHM}$ in the vicinity of the galaxy were computed. We then calculated the brightest point source that we would fail to detect at the $3\sigma$ level in an aperture of size 1.0 $b_{\rm FWHM}$ given $\sigma_{\rm rms}$ and $b_{\rm FWHM}$ as computed above. The total magnitudes we compute for such sources give the following limits for the magnitude of the SN: $m_{K'} > 19.83$ (March 1995), $m_V > 23.55$ (August 1995), $m_R > 24.42$ (August 1995), $m_I > 22.77$ (August 1995).

## 3. Discussion

Supernovae typically have peak absolute blue magnitudes of $-17 > M_B > -19$. The value of $m_B = 18.83$ that we measure in the May image is therefore what one might expect if the point source that we see is a SN in IRAS 12112+0305 being observed near its peak (the distance modulus of IRAS 12112+0305 is 37.36 so the absolute magnitude is $M_B = -18.57$, correcting for $0.04^m$ of Galactic extinction). That it is observed in a galaxy where there is independent evidence (from the high $L_{\rm IR}$ and the high CO luminosity) for vigorous star formation in progress is further suggestive that we are observing a SN. However, we do not have a spectrum of the candidate SN and so cannot formally rule out this as being some foreground Galactic cataclysmic event. It might even be a distant and compact minor planet or asteroid, as IRAS 12112+0305 lies near the ecliptic. However, given the absolute magnitude of the object were it to lie in IRAS 12112+0305, and the duration of the event, the SN interpretation is overwhelmingly the most likely one.

It is however not possible to identify the type of SN based on this measurement from the May image alone. Type Ia SN have peak magnitudes $M_B = -18.8$ ($H_0 = 75$ km s$^{-1}$ Mpc$^{-1}$) with scatter of about $0.3^m$ (Kirshner 1990). Type II SN typically have peak magnitudes $1.5^m$ fainter than this, but the scatter in this peak magnitude is large, and the brighest SN II known have peak magnitudes comparable to those of SN Ia (Kirshner 1990). Therefore this measurement by itself does not distinguish between these two cases. Furthermore, we cannot be certain that we are observing the SN exactly at its peak magnitude in which case we cannot compare our measurement with these numbers. However the value of $M_B = -18.57$ is sufficiently bright that we are probably observing the SN very close to the peak in the lightcurve. Type Ib SN, however, generally have peak magnitudes much fainter than what we observe (Uomoto & Kirshner 1986) and so can be ruled out. Finally, our measured magnitude might be affected by internal extinction from dust in the galaxy. Again, however, the observed magnitude of this SN is so close to the maximum observed peak magnitude for both Type I and Type II SN that this extinction is probably very small, even in the B-band. This is perhaps not too surprising as the SN is in an extension of the galaxy well away from its center.

We therefore attempt to determine the type of SN by comparing our May and August data with typical lightcurves. Figure 2 suggests that we are probably observing a SN II as the fading between our May and August observa-tions is greater than expected for a Type I. For this SN to be a typical Type I we would have to hide a point source $1.5^m$ brigher than our estimated limiting magnitude in Section 2, which we regard as unlikely. However the scatter around the curves in Fig. 2 is approximately $1^m$ (Kirshner 1990), so it is still possible that we could be observing a SN Ia, but it would then be a fairly extreme example. Note further that SN II can have colors as red as $B - V = 1.0$ at 90 days past maximum (Younger & van den Bergh 1985). If this object is that red, then the limit shown in Figure 2 is moved down 0.6 mag and the SN II interpretation also becomes dubious.

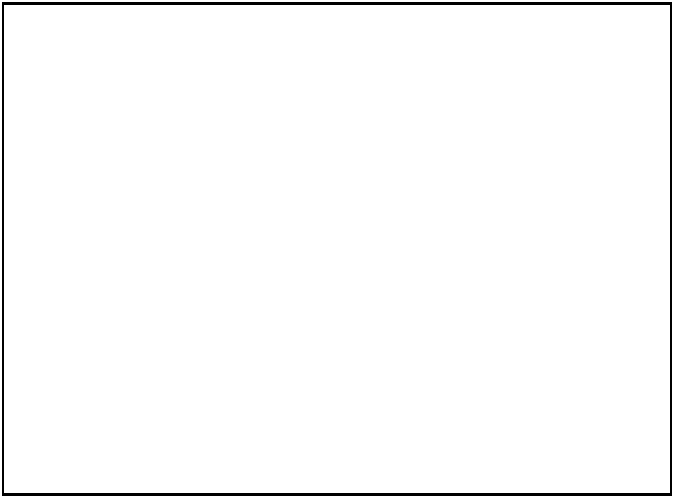

**Fig. 2.** A comparison of our data with SN lightcurves. The lightcurves are those computed by Doggett & Branch (1985) for typical Type I, Type II-P (Plateau), and Type II-L (Linear) SN. Typical scatter of individual SN around these curves is about 1 mag (Kirshner 1990). The point at the origin is our B-band point (May 1995) and the limit at 92 days is from our V-band limit (August 1995). These points were calculated as described in Section II, and plotted **assuming** that we are observing the SN at peak in May. When comparing the data and the curves, any lateral translation of the points is therefore possible. The limit assumes $B - V > 0.4$ at 92 days after peak for both Type I and Type II SN, which we regard as being conservative given the compilation of Younger & van den Bergh (1985). We ignore reddening due to dust within IRAS 12112+0305 because the internal extinction is known to be small from the observed peak magnitude of the SN. We do not plot the March $K'$-band limit, due to the lack of any comparison data from well-studied SN.

In summary, our results suggest that if we have seen a SN, it is probably a SN II, which is what is expected given the environment, in particular the high gas density and huge infrared luminosity of IRAS 12112+0305, which together are suggestive of a high star formation rate. But we do stress that this interpretation is based on a single detection, and so any conclusions drawn from it must be made with caution.

## References

Carico D.P., Sanders D. B., Soifer B. T., Matthews K., Neugebauer G., 1990, AJ 100, 70
Clements D. L., Sutherland W. J., McMahon R. G., Saunders W., 1996, MNRAS 279, 477
Doggett J. B., Branch D., 1985, AJ 90, 2303
Kirshner R. P., 1990, in: Supernova, ed. A. G. Petschek, Spring-Verlag, New York, p. 59
Landolt A. U., 1992, AJ 104, 340
Sanders D. B., Mirabel I. F., 1996, ARAA, in press
Sanders D. B., Soifer B. T., Elias J. H., Madore B. F., Matthews K., Neugebauer G., Scoville N. Z., 1988, ApJ 325, 74
Solomon P. M., Downes D., Radford S. J. E., 1992, ApJ 387, L55
Steston P. B., 1987, PASP 99, 191
Treffers R., Leibundgut B., Filippenko A. V., Richmond M. W., 1993, IAU Circ. 5718
Trentham N. et al. 1996, in preparation
Uomoto A. K., Kirshner R. P., 1986, ApJ, 308, 685
van Buren D., Jarrett T., Tereby S., Beichman C., 1994, IAU Circ. 5960
Wainscoat R. J., Cowie L. L., 1992, AJ 103, 332
Younger P. F., van den Bergh S., 1985, A&AS 61, 365